# Invention, Innovation, and Commercialisation in British Biophysics

Jack Shepherd[1] and Mark C Leake[1,2]

[1] School of Physics, Engineering and Technology; [2] Department of Biology; University of York, York YO10 5DD, UK.

## Abstract

British biophysics has a tradition of scientific invention and innovation, resulting in new technologies transforming biological insight, such as rapid DNA sequencing, super-resolution and label-free microscopy, high-throughput and single-molecule bio-sensing, and bio-inspired synthetic materials. Some advances were established through democratised platforms and many have biomedical success, a key example involving the SARS-CoV-2 spike protein during the COVID-19 pandemic. Here, three UK labs made crucial contributions revealing how the spike protein targets human cells, and how therapies of vaccines and neutralizing nanobodies work, enabled largely through biophysical innovations of cryo-electron microscopy. Here, we discuss leading-edge innovations which resulted from discovery-led British 'Physics of Life' research (capturing blends of physical-life sciences research in the UK including biophysics and biological physics) and have matured into wide-reaching sustainable commercial ventures enabling translational impact. We describe the biophysical science which led to these academic spinouts, presenting the scientific questions that were addressed through innovating new techniques and approaches. We consider these examples through the lens of opportunities and challenges for academic biophysics research in partnership with British industry. We highlight how commercial breakthroughs have emerged organically from fundamental research rather than from technology-first approaches but also discuss lessons to learn from past failures. Finally, we propose recommendations concerning future resourcing and structuring of UK biophysics research and the training and support of its researchers to ensure that UK plc punches above its weight in biophysics innovation, and a need to educate the policymakers and public that an absence of basic science impoverishes innovation.

## An introduction to the path towards commercialisation for British Biophysics

Since the inception of modern biophysics in the 1950s, biophysics researchers in UK have been at the forefront of development of new physical science technologies to address challenging, open biological questions. These biophysical tool developments have often gone hand-in-hand with progress in new mathematical and, later, computational approaches to extract, analyse and model or even simulate data acquired from experiments. Examples include X-ray crystallography and nuclear magnetic response (NMR) tools to help determine the structure of biological molecules and complexes, and measurement devices used in physiology such as rapid and precise quantification of the electrical signals propagated in nerve fibres. Traditionally, the UK biophysics community has done less well than many international counterparts in developing commercial applications of discovery-led findings stemming from academic research. However, in the past few decades this has begun to change, and in the UK several promising start-ups and spinouts have emerged. .

Here, we probe  recent case studies using testimonials from the academic researchers themselves which have been gathered as part of a Roadmap 2025 consultation exercise invited by UK Research and Innovation (UKRI) covering the 'Physics of Life' (PoL) research activities and community in the UK[1], including pivotal innovations in DNA sequencing, new detection and biosensing technologies, and the development of new biologically inspired materials. We articulate the basic scientific questions and challenges in each case and discuss the biophysics approaches developed to help address these. We then describe the timeline of the development of product to market for each, from the words of the developers themselves, covering aspects of the relevance of original discovery-driven research at the physical-life sciences interface in catalysing industrial impact, the challenges *en route* and the existence of enabling structures and resourcing. We then discuss some of the wider aspects of the more general commercial value of continuing to support discovery research in areas relating to biophysics, the role of early career researcher training and support in the UK biophysics, and speculate on the opportunities for improving the pipelines to ultimate commercialisation by considering potential new structures for the biophysics research community in the UK to help researchers to join the dots between discovery research and translational impact.

## Case studies from academics in the UK who have spun out biophysics related discovery research into commercialised innovation

*Carrie Ambler*

Carrie Ambler (Fig. 1) is a Professor of Biosciences in Durham University and Fellow of the Wolfson Research Institute for Health and Wellbeing. In 2016 she joined up with two partners to spin off LightOx in England, founded on core interdisciplinary principles and practices between the life and physical sciences interface. LightOx comprises teams from Durham, York and Oxford universities, and is a leading biotech aimed at developing new therapies for oral cancers.

Attempting to discover the mechanisms of the immune response in repairing skin damage[2] was the initial curiosity-driven stimulus that led to general insights about how reaction oxygen species (ROS) elicit cell damage. These investigations led Ambler and team towards investigating a chemical known to be a photosensitiser that could be utilised to elicit the production of ROS upon being activated by ultraviolet light[3]. It was only at this stage following years of basic scientific discovery that Ambler pivoted towards the prospect of developing smart therapeutic chemicals that can controllably destroy epithelial cancer cells.  The key biophysics innovations of the LightOx technology comprise the use of photophysical activation chemical reagents which transform from an inactive to activated state. LightOx pioneered the development of new light-activated therapeutics which upon activation are cytotoxic and so have primary applications in targeting relatively physically accessible cancers, in particular oral cancers, for example a condition called 'Oral Epithelial Dysplasia', a pre-cancerous condition presenting with patches on the lining of the mouth. New light-activated drugs developed by LightOX can be applied to these patches then activated by light to precisely destroy the epithelial dysplasia and thus stop progression to actual cancer and associated costly and onerous surgery subsequently.

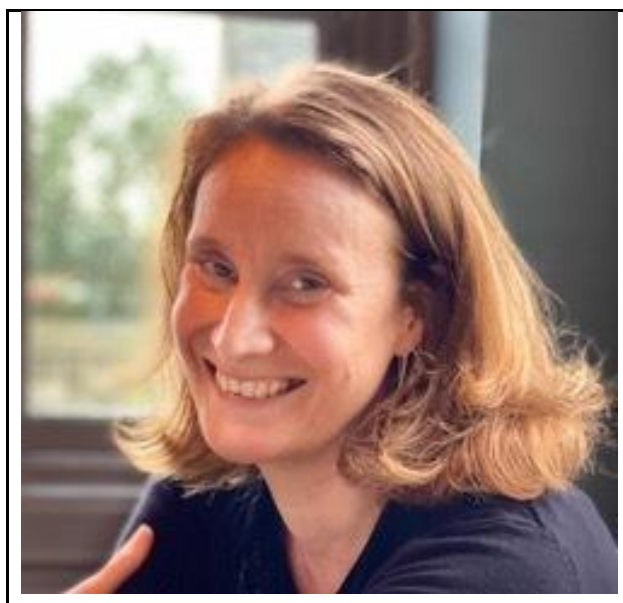

**Carrie Ambler, Durham University:** *"We've been able to leverage this interdisciplinary work for a good amount of industrial funding, both in terms of LightOx supported funding, but importantly, we've, received nearly £2M of funding from Innovate UK"*

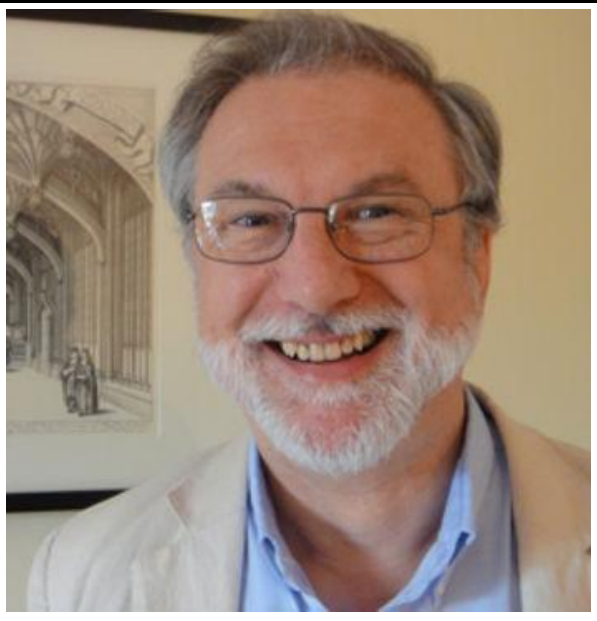

**Hagan Bayley, University of Oxford:** *"The multidisciplinary approach can be highly profitable and indeed led to the formation of Oxford Nanopore in 2005"*

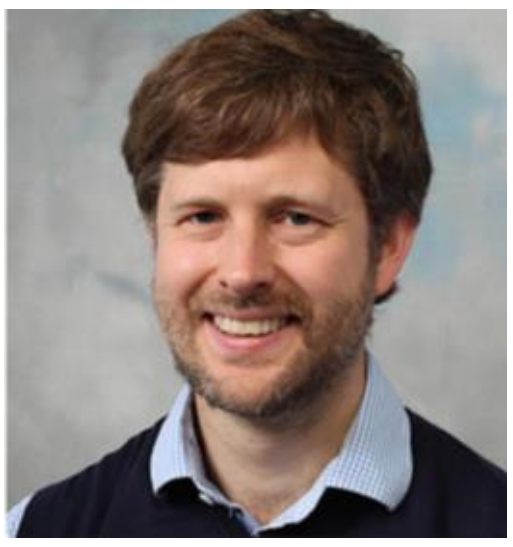

**Timothy Craggs, University of Sheffield:** *"The success of this current recruitment drive relies on a strong pool of well trained, multidisciplinary scientists, an area in which the UK has historically been strong - but must maintain for the good of many emerging companies in this space"*

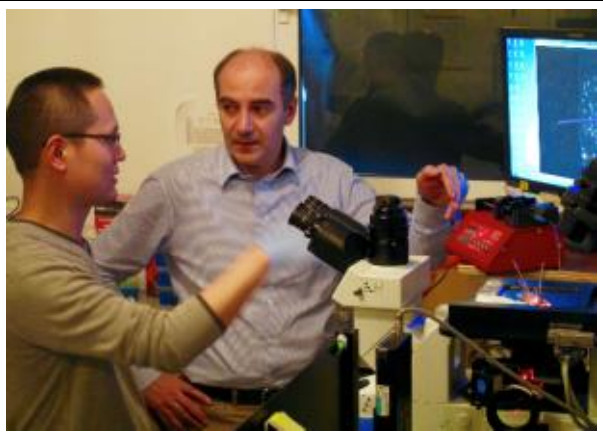

**Achillefs Kapanidis, University of Oxford:** *"These technical innovations, which were initially exclusively driven by curiosity-drive, basic-science efforts, led to the development of tools and knowhow… and created an appetite for biological scientists to access the exciting new instruments"*

**Figure 1.** Testimonials from biophysics entrepreneurs Carrie Ambler, Hagan Bayley, Timothy Craggs and Achillefs Kapanidis.

Ambler built the company from the ground up, including strategic vision, mission and values, setting up and residing on board, ensuring robust governance based on strong ethics and transparency. She was instrumental in securing funds, including a critical early income stream through a life sciences reagents business and multi £m capital from individual investors and venture capital. Headed up the design and development of the LightOx PhotoReact 365(TM), and established a domestic and international network to advance LightOx's R&D.

In the PoL Roadmap, Ambler says: "In all ways, at this point, my both academic research as well as my company research is interdisciplinary. The essential essence of what we have been doing is been taking chemicals that have been generated with novel light activated or photo physical properties, and we've been using them as diagnostic tools, as research reagents for other academics, but more importantly, in terms of what LightOx does, has been using these light activated chemicals to build new therapeutics for treatment of early stage precancerous lesions in the mouth, where there is no current decent therapeutics available, and also as an antimicrobial treatment for people with chronic, antibiotic resistant bacterial infections. So how has interdisciplinary research helped us? It's in every way we have the essential need to build new chemical molecules that have innate, light activated physical properties. But to be able to maximize our ability to use those, we need to understand the underlying photophysics, things like lifetime measurements. We understand how the electronic states work, and even in terms of the therapeutics, be able to understand what, what type of photosynthesisers we've generated, and this work has been essential for, again, being able to select the right therapeutics to go forward, because we need to understand the mechanism actions behind that. So in my academic work, I've worked really closely with John Girkin [Dept of Physics, Durham University] , and we have just graduated our first PhD student together jointly, who has spent quite a long time looking at things like fluorescent lifetime, and we're looking at things like phlegm measurements and being able to look at how our chemicals dock within specific biological sites using these photo physical properties. The student is now going on to work with Helen fielding at UCL again, taking it that next stage of the physical chemistry aspect of these molecules to continue to understand how the energetic states work with this chemistry.

"So, in terms of the impact this has, obviously, it has impact on being able to do drug selection. We've been able to leverage this interdisciplinary work for a good amount of industrial funding, both in terms of LightOx supported funding, but importantly, we've, received nearly £2m of funding from Innovate UK in the past year as well."

*Hagan Bayley FRS, FLSW*

Hagan Bayley (Fig. 1) was recognised by the Science Council in 2014 as "one of the UK's 100 leading practising scientists" and holds the position of Professor of Chemical Biology at the University of Oxford. He applied biophysics with protein and organic chemistry to the develop protein pores as biosensors and "nanoreactors", leading to the formation of DNA sequencing company Oxford Nanopore.

The development of the basic biophysics of nanopore-based sensing originated around researching a natural bacterial toxin called alpha-hemolysin (aHL), which is excreted by *Staphylococcus aureus* bacteria to help out-compete microbes from other species in the immediate environment. The aHL protein is secreted as the monomer but then undergoes self-assembly upon binding to a target cell membrane to form a nanopore with 7 subunits whose physiological role is to eradicate the transcellular protonmotive force by enabling free diffusion of protons across the membrane through the nanopore, thereby ultimately killing the cell. Seminal biophysics development explored the use of correlative technologies to measure the aHL assembly process in fluorescence using total international fluorescence (TIRF) microscopy with simultaneous ion conductance measurements across artificial lipid bilayers through the emerging nanopore. Bayley and team later discovered that by using an appropriate spatial 'adapter' molecule of aminocyclodextrin the level of nanopore current detected became sensitive to whether DNA was being translocated through the nanopore. A revelational discovery was that each of the four DNA nucleotide bases would occlude the nanopore during translocation in slightly different ways such that the specific level of drop of ion flux could be correlated directly to the type of nucleotide base – hence, an entirely new approach to sequence DNA was born[4].

Bayley indicates: "My experience with nanopore sensing and sequencing, beginning with the detection of metal ions as far back as 1990, has reinforced my belief that strong fundamental science at universities or research institutes almost inevitably precedes commercialisation. Our initial question was of completely academic interest: "How can a water-soluble bacterial protein toxin, staphylococcal alpha-haemolysin, assemble into a transmembrane pore". The beautiful X-ray structure of the heptameric pore from 1996, which resulted from that work, placed nanopore sensing on a rational basis and future progress in that area required the similar underlying support of fundamental studies. Alpha-haemolysin was the original nanopore sensor. Further work refining nanopore sensing and sequencing was highly interdisciplinary and encompassed several subdisciplines in the physical and biological sciences, which have become critical to the Physics of Life community. In research, it has been essential to breach the boundaries that have been used to establish university departments or rationalise undergraduate teaching. My own laboratory in a Chemistry Department has employed: single-molecule biophysics notably electrical recording, 3D printing and microfluidics, surface and lipid chemistry, organic chemical synthesis, a wide variety of protein engineering techniques, optical microscopy, mammalian cell culture, and more. While we are not the top experts in each of these fields, it is clear nonetheless that the multidisciplinary approach can be highly profitable and indeed led to the formation of Oxford Nanopore in 2005. Since then, academic research findings continue to be fed into the company."

*Timothy Craggs*
Tim Craggs is a Senior Lecturer at the University of Sheffield and Founder of Exciting Instruments.

The core Exciting Instruments technology emerged from curiosity-driven biophysics research performed by Tim Craggs (Fig. 1) as a postdoc during his time in the Clarendon Laboratory of the University of Oxford during which he was investigating how to use Förster Resonance Energy Transfer (FRET) at a single-molecule level to enable high-precision quantification of protein complex structural conformational transitions[5]. FRET is a non-radiative energy transfer that results from the resonance coupling between electric dipoles associated with donor and acceptor pairs of fluorophores. This diploe coupling occurs over a length scale of ~2 to 10 nm and so is particularly valuable as a metric of putative interaction between the molecules to which the donor and acceptor molecules have been tagged. The fluorescence excitation and detection technology was based around using two colour visible light lasers focused down to coincident confocal volumes. Sample illumination by the two colours could then be alternated at high speed to enable rapid stroboscopic excitation of either the acceptor or donor molecules thereby minimising optical blead-through between the two fluorescence emission detection channels which had been a technical limitation previously. Rather than illuminate a whole microscopy field of view, the confocal volumes are positioned deep in a sample of purified proteins and capture the FRET behaviour of molecules which happen to diffuse through the illuminated volume. This approach was used by Tim Craggs with others to investigate the structural dynamics of the DNA polymerase enzyme and its ability to reject incorrect nucleotide bases in a DNA sequence being replicated[6]. Sensible labelling locations for these FRET dyes on the DNA polymerase complexes enabled a picture to be built up for open-close transitions during the few milliseconds of duration in which the molecule was excited by the laser before diffusing out again. Craggs and team then worked over the next few years to develop this technology into an open-source platform that could reliably quantify single-molecule FRET, and it was this that led to the first standalone commercial platform called the smfBox[7].

Craggs says: "Necessity and multidisciplinary science — the mother of invention! Starting out as a new PI in Sheffield in 2016 I was keen to crack on with an ambitious research program, interrogating conformation and dynamics of DNA - at the single molecule level. I had developed a strong international reputation in this field but was faced with a significant problem. I needed an instrument to perform my single-molecule measurements, and (as generous as Sheffield were) the UK uni start up package was nowhere near the level needed to buy a commercial instrument. So, with the help of two dual honours physics - chemistry integrated Master's students, I set about building my own instrument for a fraction of the cost of available commercial versions. Building this instrument, the smfBox, required a unique blend of skills and knowledge, spanning optical physics, electronics, coding and biology - as this is where all the actual applications for the instrument lie. This multidisciplinary approach was essential to success of the project, which formed the basis of the first product for our spin out company Exciting Instruments. We quickly realised that other labs would benefit from our instrument, but they didn't have the right mix of skills to build it themselves. So, we formed the company Exciting Instruments, to take the smfBox and turn it into the EI-FLEX, a benchtop instrument for single-molecule FRET and FCS. The FLEX takes all the physics from the smfBox but presents it in a way that is useable even by undergraduate biologists. It is this ease of use, driven by the deep knowledge of both physics and biology, that has truly allowed us to democratise single molecule methods for the widest possible user base. Putting complex physics and optics into the hands of biologists, through simple to use software and analysis, has opened up the market for our instruments, enabling us to achieve slaws revenue of >£1.6M since spinning out the company in September 2021. We are now expanding the Exciting Instruments team, with all our staff needing to communicate across the disciplines of physics, chemistry and biology. The success of this current recruitment drive relies on a strong pool of well trained, multidisciplinary scientists, an area in which the UK has historically been strong - but must maintain for the good of many emerging companies in this space."

*Achillefs Kapanidis*

Achillefs Kapanidis is a Professor of Biological Physics at the University of Oxford and a board director and consultant for the biotech company Oxford Nanoimaging.

Prior to establishing Oxford Nanoimaging, Achillefs Kapanidis (Fig. 1) was involved in several discovery-led biophysics research projects which focused on the application of bespoke single-molecule fluorescence microscopy tools and approaches for understanding structural dynamics of single-molecules, including helping to develop and apply alternating laser excitation and FRET based tools to investigate how the genetic code is read out by RNA polymerase by utilising the stored elastic energy within "scrunched" DNA[8].

Kapanidis writes: "The invention of single-molecule fluorescence methods revolutionized biophysics and molecular biology by breaking the diffraction limit in microscopy, offering real-time views of biochemical reactions and enabling ultrasensitive detection. When I joined Oxford (2005), my group started building several customised microscopes for single-molecule detection in order to perform sophisticated biophysical measurements of biological mechanisms, such as bacterial gene transcription and DNA repair. The complexity of the questions and the need to collect sufficient statistics to support the ground-breaking basic-science observations required that the instruments are stable, robust, streamlined, and practical to use; further, the instruments needed support by versatile, efficient, and stable software, both for data acquisition and data analysis.

"These technical innovations, which were initially exclusively driven by curiosity-drive, basic-science efforts, led to the development of tools and knowhow that led to two main outcomes: first, the publication of the novel results from the new instruments raised further the profile of single-molecule method outside the realm of biophysicists, and created an appetite for biological scientists to access the exciting new instruments. Second, the strength of the group in building instruments and software for data acquisition/analysis provided confidence for the group to engage in microscope miniaturisation projects. One such project led to the development of the Nanoimager, a robust miniaturised fluorescence microscope that excels in single-molecule and super-resolution imaging and replaces unstable and cumbersome custom-built microscopes which used to occupy entire rooms and requiring laser interlocks and massive optical tables to operate. The Nanoimager was commercialised by Oxford Nanoimaging, an Oxford spin-out that I co-founded in 2016."

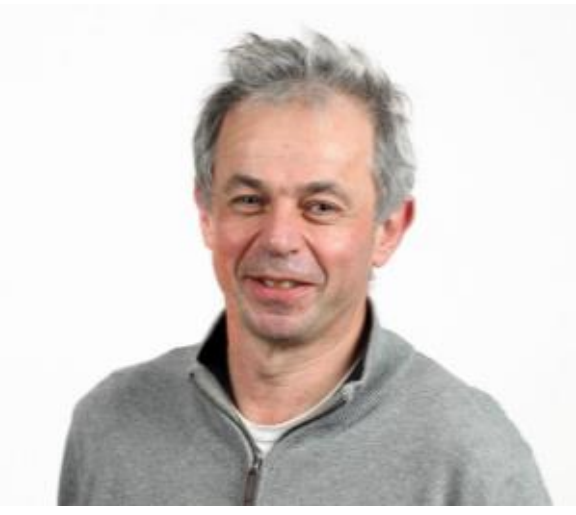

**Sir David Klenerman, University of Cambridge:**

*“The important element behind our success was the funding of multi-disciplinary basic science to explore the frontier between disciplines, especially cutting-edge physical science… and biology”*

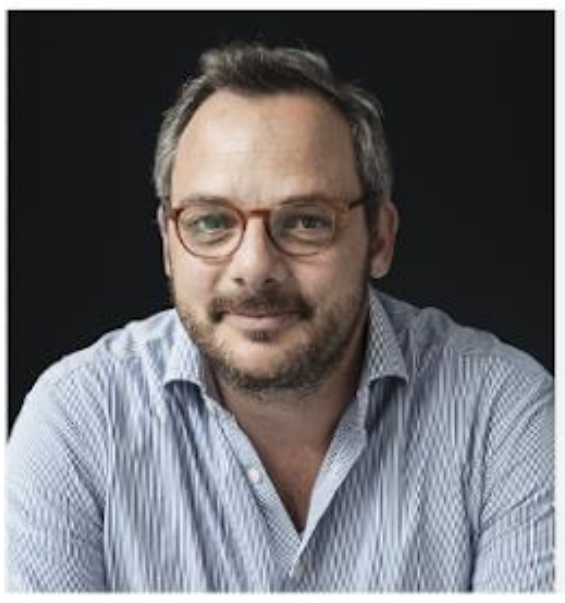

**Phillipp Kukura, University of Oxford:** *“It cannot be emphasised enough how important supporting basic research has been”*

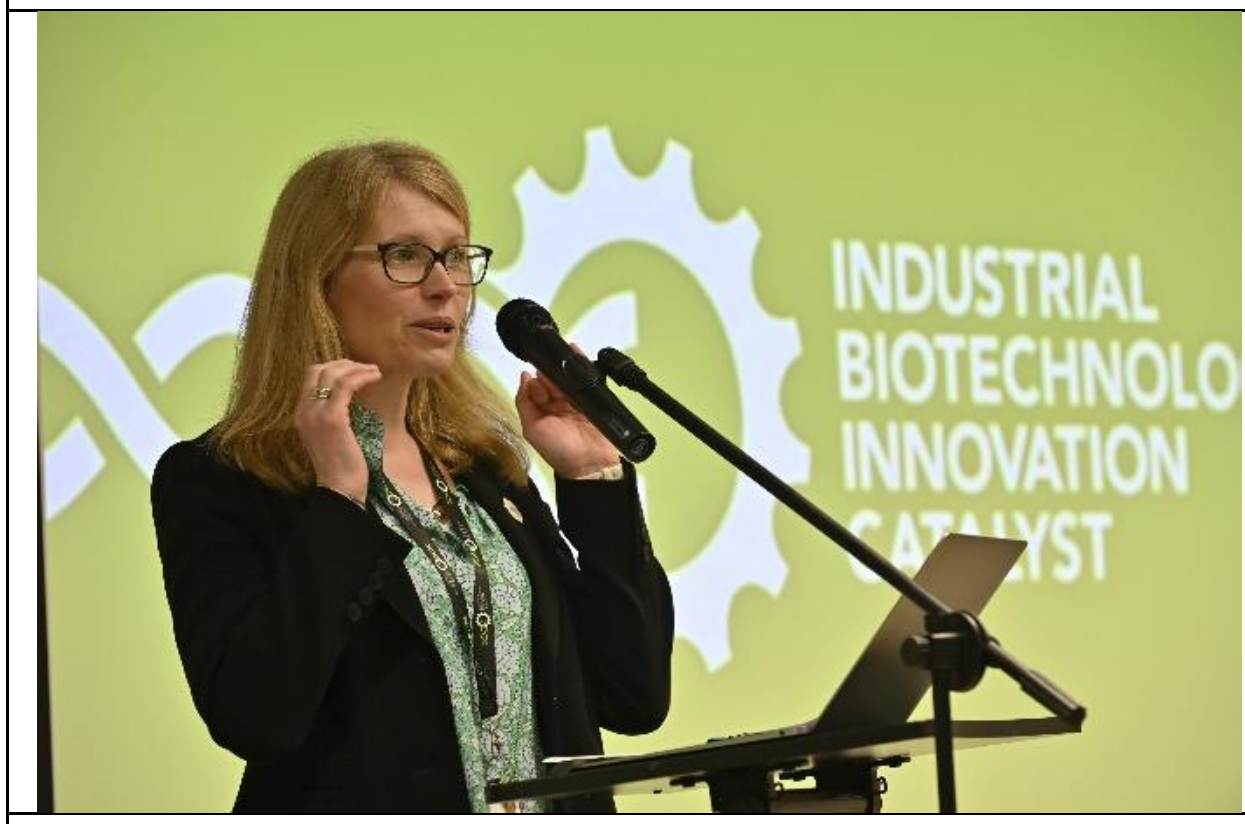

**Aline Miller, University of Manchester:**

*“Novel materials can provide biocompatible, injectable, and translatable scaffolds for in vivo applications such as drug and/or cell delivery for tissue repair and regeneration”*

**Figure 2.** Testimonials from biophysics entrepreneurs Sir David Klenerman, Phillip Kukura, and Aline Miller.

*Sir David Klenerman FRS FMedSci*

Sir David Klenerman (Fig. 2) is a professor of Biophysical Chemistry at the University of Cambridge. In 2007 the British owned company Solexa which he jointly created was acquired by the US company Illumina, which rolled out the DNA sequencing pioneered by Solexa on an extensive global scale.

The scientific principles of Illumina’s next-generation DNA sequencing technology were established in the early 2000s in biophysics experiments in the University of Cambridge focused around ultrasensitive coincidence fluorescence detection of single DNA molecules – in essence making basic science discoveries about the effect that labelling DNA with more than one colour of fluorescent dye can make on improving the single-molecule detection efficiency compared with just using single colour fluorescence detection[9].

Klenerman says: “With my colleague Shankar Balasubramanian I co-invented next generation DNA sequencing which enabled sequencing speeds and costs to be reduced by a factor of about a million allowing the genomes of an individual person to be sequenced at scale. To date 90 % of the DNA sequence data in the world has been taken on this platform and it is starting to be used routinely in hospitals for prenatal screening, diagnosis of rare diseases and cancer diagnosis and treatment. The important point is that the idea that underpins next generation DNA sequencing came from a blue sky research project funded by the BBSRC to observe DNA polymerase copying DNA in real-time, exploiting new single molecule methods to watch biology in action. DNA sequencing as a potential application is not mentioned anywhere in the grant, however by performing this cutting-edge science we realised we could repurpose the experiment for rapid DNA sequencing so formed a company, Solexa to do this.

"The important element behind our success was the funding of multi-disciplinary basic science to explore the frontier between disciplines, especially cutting-edge physical science (in this case single molecule fluorescence) and biology. Basic research should be funded without expectations of quick results, as history shows that incredible breakthroughs happen when the right elements are brought together. The experience of the development of COVID-19 vaccines and base editing are other examples. Secondly there are tremendous opportunities in the application of physical methods to biological problems enabling new measurements to be made that were simply not possible before. Super-resolution microscopy and next generation DNA sequencing are two recent examples. To encourage work at this interface funding is an essential requisite. It also requires physical scientists to find out about appropriate biological problems and the biologists/biomedics to find out what might be possible and then enable pilot experiments to explore the area and see if there is a good way through. Ironically, our specialisation at University and Departmental structure creates barriers to doing this effectively. It also requires a change in culture where working together in a multi-disciplinary team to do impactful research between disciplines is seen to be as important as doing quality research by oneself, something that is common in industry but still rarer in academia."

*Phillip Kukura*

Phillip Kukura (Fig. 2) is a Professor of Biophysical Chemistry at the University of Oxford and former CEO of the company he co-founded, Refeyn.

Over a decade ago, Phillip Kukura's Oxford-based team reported on the new potential for developing "mass photometry", i.e., the application of interferometric scattering mass spectrometry (labelled initially as iSCAT[10] and later as iSCAM[11] to discriminate between relatively small differences in total mass of a range of calibration proteins manifested as surface-immobilised biomolecular complexes *in vitro.* Ultimately, these became sensitive enough to discriminate between different lower-order oligomerisation states (monomer-tetramer) of purified BSA, as well as larger mesoscale structures such as biomolecular filaments. iSCAM/iSCAT has high enough contrast to enable imaging of single protein molecules without requiring a fluorescent label. The sample is localized on a microscope coverslip at a glass/water interface illuminated using coherent laser light. Light intensity $I_d$ detected is the sum of reflected light from this interface and that scattered from the biomolecules on the surface:

$$I_d = |E_{\mathrm{ref}^2} + E_{\mathrm{scat}^2}| = |E_{i^2}|(R^2 + |s|^2 - 2R|s|\sin\varphi)$$

$E_i$, $E_{ref}$ and $E_{scat}$ are incident, reflected and scattered light $E$-field amplitudes, $R$ and $s$ the reflected and scattering amplitudes and $\phi$ the phase between scattered and reflected light. For small length scale scattering objects, the $|s|^2$ is close to zero because the Rayleigh scattering cross-section, and hence the scattering amplitude $|s|^2$, scales with $V^2$ for a small scattering particle whose radius is much less than the wavelength of light. However, the interference term which is $2R|s|\sin\phi$ only scales with $\sim V$ and so is far less sensitive to changes in scatterer size, and it is this term which is the physical basis of iSCAT/iSCAMS. An iSCAT/iSCAM microscope comprises a confocal laser illumination volume laterally scanned across the sample, though instead of detecting fluorescence emission the interference term is extracted by combining a quarter wave plate with a polarizing beamsplitter; this utilizes the phase difference between the interference term with respect to the incident illumination, and rotates this phase to enable highly efficient reflection of just this component at the polarizing beamsplitter, which is directed not through a pinhole as for the case of traditional confocal microscopy but rather onto a fast CCD camera.

Kukuka comments: "The development of mass photometry is a case in point when it comes to the importance of fundamental research at the interface of the physical and life sciences. The original research paper describing the technology, which was entirely novel at the time, was published in April 2024. Refeyn, a spin-out from the University of Oxford was founded in June 2018, and the first instrument shipped to customers in October 2018. The company has since grown to 180 employees worldwide, has shipped more than 300 instruments and has been used (not cited!) in over 500 scientific publications. More than 50% of sales are in the biopharma sector, in applications ranging from drug development, to production, to R&D. Mass photometry, and the associated products from Refeyn are playing a central role in enabling the next generation of therapeutics, in particular cell and gene therapy and mRNA delivery. The company has won a number of national and international awards for its technology and is establishing mass photometry as one of the key future bioanalytical technologies. It is also one of very few technologies emerging from single molecule optics that has been successfully commercialised and is being broadly adopted. It cannot be emphasised enough how important supporting basic research has been in this regard. Mass photometry is based on detecting single biomolecules in solution using light scattering alone. Prior to 2014, this was considered completely impossible. The concept of single molecule mass measurement by light was completely novel in 2018 - this was an application that nobody had expected, including those developing the technology. It emerged from the interface of high-performance optics in the Kukura group and native mass spectrometry applied to proteins in the Benesch, both in Oxford. The underlying research in interference microscopy had been funded largely by the ERC as a high-risk project - with no connection to single molecules or mass measurement whatsoever. In fact, any grant proposal claiming such capabilities would have most certainly been rejected due to a lack of feasibility. The key aspect to success here has been the support of cutting-edge research at the interface of physics and biology. Quantitative measurement is becoming ever more important in the life sciences and the broader health sector, enabling new avenues to intervention. The underpinning advances can only come from breakthroughs in Physics but co-developed with life scientists. For mass photometry, it was the back and forth between what the technology could do, and what genuinely novel information could be provided - and used - by the user, that drove the development. Supporting research at this interface, while asking the key question - why? - is a key way to ensure future breakthroughs and competitiveness for the UK in this enormously important future growth area."

*Aline Miller*

Aline Miller (Fig. 2) is a Professor of Biomolecular Engineering in the School of Chemical Engineering and Analytical Science at the University of Manchester. She has authored in excess of 100 research articles and owns five patents, in addition to being the School's Research Director and leading Gender Equality and Athena SWAN activities within the University. She has won several awards including the Exxon Mobil Teaching Fellowship in 2004 and in 2008 was awarded The Royal Society of Chemistry MacroGroup UK Young Researchers Medal and also The Institute of Physics, Polymer Physics Group Young Researchers Lecture Award for her work on self-assembling materials. More recently she won the 2014 Philip Leverhulme Prize for Engineering and was shortlisted for the 2014 WISE Research Award. In the same year she co-founded her own spinout PeptiGelDesign Ltd.

For over half a century, the biomaterial properties of hydrogels have attracted significant interest for a range of biomedical devices and applications, such as use as scaffolds in tissue engineering and for controllable drug delivery vehicles. Aline Miller made early discovery-led progress in establishing the utility of the model protein hen egg white lysozyme to explore self-assembly and gel effects and using a range biophysics tools to characterise these including environmental scanning electron microscopy (ESEM), cryo-transmission electron microscopy (cryo-TEM), rheological measurements and Fourier transform infra-red spectroscopy[12].

In Miller's words: "Driven by the question of what molecular parameters drive a protein to fold into specific 3-dimensional structures, we started a blue-sky research project with a PhD student and some start-up funds from the University, and 20 years later ended up founding, growing and exiting a start-up company, Manchester BIOGEL Ltd. The technology developed in Manchester is based on understanding and controlling the self-assembly of small peptides across the length scales, to design bespoke hydrogel materials. We switched from proteins to peptides to give us reproducibility in self-assembling pathway and our toolbox is the pool of 20 natural amino acids that make up natural proteins. We combine these amino acids in different sequences dialling in electrostatic, hydrogen, hydrophobic, pi-pi stacking and van der Walls interactions and over the years have developed a platform technology where we can control material property and function by simply by designing the peptide molecule primary sequence.

"The platform allows the engineering of ethical, animal free materials with properties that can be tailored to mimic the 3-dimensional micro-environment, cell niche, in which cells live. Cells have very different requirements depending on their origin, nature and function, and this technology enables the design of scaffolds with properties and functionality customised to each cell type needs. Consequently, this work opens-up new possibilities in the biomedical field: These novel materials can provide biocompatible, injectable, and translatable scaffolds for *in vivo* applications such as drug and/or cell delivery for tissue repair and regeneration; or fully synthetic and controlled micro-environments for in vitro applications such as growth of multi-cellular organoids for disease modelling and drug toxicity and efficacy testing.

"Demand for these hydrogel materials from collaborators led to the company Manchester BIOGEL (formerly PeptiGel Design) being formed in 2013. Over the years I led the company as CEO and raised £4M+ in investment from Innovate UK, Venture Capital, Private and Catapult Venture Funds, grew to a team of 10, scale hydrogel manufacture and the company became revenue generating. This led to Manchester BIOGEL being listed as one of the Top 10 BioTech Start-Ups in Europe by Start-Up City in 2021, winning Best New Life Science Product 2021 and I navigated a successful exit, with the company technology being sold onto Cell Guidance Systems in 2023."

**Emerging findings relating to the success stories of commercialisation of UK biophysics research**

1. *Serendipitous curiosity-driven findings can motivate translation, but transformative impacts which do not involve initial open scientific curiosity are rarer.*

A common feature of the recent UK biophysics commercialisation successes outlined above, is that the preliminary development of the new approach in biophysics was coupled to basic, discovery-led research, as opposed to involving the developments of new technologies with a subsequent stage of looking for potential applications of the new technology. In other words, a new approach or technology was not created in isolation – a curiosity-driven research question motivated its development. This is an intriguing and arguably counter-intuitive observation should be tensioned against common practices seen in many UK universities, and also encouraged by research funding bodies, to resource 'sandpits' in which typically attempts are made to matchmake the 'problem-owners' (i.e. the researchers with primary interests in addressing specific open scientific questions) with appropriate 'solution-providers' (i.e. typically researchers who have developed a new technology and/or analytical approach). A point to reflect on here might be where the balance between resourcing basic discovery research versus encouraging technologists to essentially 'fish' for applications should best lie if one of the ultimate aims is to maximise translational impact towards innovation and commercialisation.

2. *Approaches that integrate multiple disciplines at the physical-life sciences interface beyond just physics and biology can enable unprecedented new understanding which can lead to disruptive new biophysics technologies.*

Several of the UK entrepreneur case studies that we have detailed above hammer home a message that is the close integration of physical and life sciences that has been absolutely pivotal towards enabling new scientific insights in the early stages of discovery research that ultimately led towards a commercial product. An interesting feature of this finding is that this goes beyond just the interface between physics and biology, but very much extends also into areas of chemistry, engineering and materials science. Similarly, several of the entrepreneurs themselves, although making seminal developments associated with biophysics, have been hosted by academic departments which are not simply 'core' physics or biology. In this regard, there are many examples of biophysics innovation that have benefitted from expertise which extends beyond just the semipermeable membrane between physics and biology.

3. *UK biophysics spinouts exhibit geographical diversity beyond the Golden Triangle despite the majority of research investment being focused on the Golden Triangle.*

Traditionally, there has been a perception that the core of UK research activity is concentrated in the 'Greater South East' region of England (sometimes referred to as the 'Golden Triangle' of Cambridge, Oxford and London.) However, recent data collated from research income and impact of publication outputs suggest that there is a not a strong correlation between the two[13]. The relative lack of proportional transformative impact emerging from the Golden Triangle is reflected in three out the seven UK biophysics commercialisation case studies above emerging from Durham, Sheffield and Manchester. Observations such as these have motivated discussions on how efficient the research investment in favour of Golden Triangle institutions is towards delivering transformative research outcomes. It is clear that there is an important level of investment in innovation centres in and around Golden Triangle institutions, but several institutions outside the Golden Triangle have research and enterprise hubs helping to catalyse the development of commercial opportunities from discovery research. There is also a strong regional stimulus in several parts of the North of England in particular in working strategically with relevant Combined Authorities to maximise engagement between academia and local industry[14].

4. *There is precedence from other areas of scientific research beyond biophysics for implementing specific resources and research structures which can help join the dots between discovery and translation.*

Many of the biophysics entrepreneurs questioned articulated the value of academic institutional pump-priming funds, often comparatively small equivalent to the order of ca. £10k funds often routed through equivalent EPSRC and BBSRC Impact Accelerators budgets which are locally administrated by academic institutions, to help road-test potentially risky research ideas with a view towards exploring commercialisation opportunities. For more advanced commercialisation ideas, the value of more extensive funding from InnovateUK was cited. One bottleneck with moving forwards with academia-industry partnerships is the negotiation of an appropriate intellectual property (IP) agreement, and there is value in using template framework agreements as an early start point towards the final aspiration of a signed, tailored IP contract that all parties are happy with.

Although there are examples of good practice within academia, currently the framework of management and resourcing within many UK universities as a whole is arguably not as efficiently structured as it could be to favour new commercial innovation. Academics are not as a rule trained to be effective managers, and it shows – from cases of inconsistent and mixed levels of support from the level of postgraduate students up to major cross-disciplinary research consortia. The prerequisites for academic promotion are arguably in some cases increasingly focused on research grant incomes and associated success rates of research fellowship applications rather than on a balanced portfolio involving experience, vision, and knowledge that relate to commercial opportunities. There are also anecdotal cases of academic departments being run in effect as individual fiefdoms, expert technical staff considered extraneous with only short-term employment prerequisite on a 'business case'. The challenge of senior researcher academics who act in effect as isolated islands inhibits the formation of 'academic continents' of research through true collaborative activities.

5. *Open democratisation of innovation that leads to a commercial product can increase overall research productivity more efficiently than isolated research teams developing in-house non-commercialised technologies focused on granular scientific questions.*

There is often a tension between patenting opportunities towards commercialisation working against the principles of openness and accessibility to the wider research community which is fundamental to many aspects of biophysics research. The key to moving forwards is perhaps to ensure that the right balance is struck. Early engagement with academic research and enterprise teams to help draft template IP frameworks and preliminary non-disclosure agreements is helpful to at least allow preliminary conversations to occur between potential industrial collaborators and investors. In many sectors of research relevant to biophysics there are good examples that more openness in research can be invaluable towards creating a future market base for a commercial product. This is seen is the bioimaging community with democratising designs of bespoke microscopy, the structural biology community which upload the coordinates of biomolecular structural datasets in a timely manner, and the underpinning biophysics computational communities in making software openly accessible, for example through GitHub, and now - almost ubiquitously - the use of preprint servers to enable early dissemination of research findings to the wider researcher community. A key is to structure IP agreements in such a way as to enable a sensitive balance to be struck to enable such openness through agreed mechanisms with fixed timelines for discussion between the academics and industrial and/or investment partners on a case-by-case basis regarding making discoveries public, whilst still retaining appropriate confidentiality of key commercially sensitive details.

*Success or failure of spinout companies may not be predictable, but institutional and research-council activities can increase the chances of success, for example facilitating tailored training for research group leaders in how to establish a new startup and maintaining "institutional knowledge" and continuity of research expertise through open contracted academic research staff.*

One way to better support and encourage problem-solving science that leads to ultimate invention and innovation is to provide more sustained support for early career researcher who are the drivers that implement the research. But a key may be to develop stable creative career pathways that are not solely focused on developing in a traditional academic sense, but rather also comprise significant non-academic opportunities. One of the most important and underrated factors here is 'institutional knowledge' – a pool of expertise comprising a continuity of researchers working in the same department who remember not only past successes and failures but the tricks and hints that are won of experience. These individuals as a rule are not the high-flying academics or researchers holding independent fellowships but are senior postdoctoral researchers and technical support staff. Typically, however, the majority are employed on relatively short-term fixed contracts associated with specific research grants and then are compelled to move to other research teams, often hosted by other instructions, at the end of each such contract. Having, instead, such people in an 'open' contracted post, for example denoted as an 'research officer', a specialist role at a more senior level than postdoc or research technician, something like a research specialist, may help to avoid losing that vital institutional knowledge every ca. 3 years of the research grants cycle. Could such specialists not only provide delocalised support to more than one research group holding a research grant but also be seconded or 'bought out' by spinout companies needed advanced technical assistance?

**Lessons to Learn from failed commercial ventures**

On the flip side of the successful commercial ventures highlighted above, however, there are many examples of failures, and lessons can be learnt from these. One good example is that of the development of magnetic resonance technologies both for high precision nuclear magnetic resonance (NMR) measurements for determining structural features of biomolecules and for medical physics imaging detector innovations including magnetic resonance imaging (MRI). Such tools have enabled drug design and variants of functional MRI with transformative diagnostic capabilities, originating in part from Peter Mansfield's work in Nottingham (Nobel Laureate, 2003) and the Oxford Enzyme Group under Rex Richards and George Radda. All drug companies use NMR, and this technology was first developed in the UK in the 1950s. In subsequent decades, the UK was preeminent in magnet technology with the companies Oxford Magnets and Magnex being at the forefront in the 1960s – 1990s, stimulated by this UK-based lead in basic discovery research. However, with no Government intervention to prevent buyouts, these companies were eventually bought out by foreign investment consortia and then moved abroad (in particular, to US, Germany and Japan). In the years since, there has been minimal magnet technology or development left in the UK.

A second example involves John B Goodenough (ex-Oxford, Nobel Laureate 2019) who developed lithium battery technology in the UK but went to the US to bring it to commercial fruition, with the UK then being left with a much-depleted discovery research base and with very limited investment in this science today. This technology underpins several fields including biophysics, and production has been overtaken by other countries.

And thirdly, despite the huge and inspirational efforts of Kate Bingham and colleagues in helping to develop COVID vaccination technology, the UK now has no vaccine mass production facilities *per se*. Those set up during the pandemic, with great UK Government announcements, are now closed. The UK is dependent, even now, on overseas production, in particular from India and the Netherlands.

A crucial lesson to learn here is that if Government support for biophysics innovation does not continue once startups are established then not only does the UK risk not reaping the rewards of its ultimate mass-scale commercialisation, but worse still a void can be created that then has a very detrimental reverse-impact on the associated area of discovery research from which the innovation originally emerged.

**The challenges facing UK biophysics commercialisation, and potential practical opportunities to address these:**

- *Can more be done by host academic institutions to train academic biophysics researchers to facilitate biophysics innovation and commercialisation?*
    - Many universities host excellent business schools with associated expertise in business management, and UK universities typically have enormous untapped reserves comprising thousands of willing biophysics early career researcher 'trainees' who are desperate for hands-on experience of interfacing with industry. Could these resources be creatively utilised to help train academics in-house to be better innovators through developing better practices of project management? be Why not ask them how best to manage a project workflow? For example, through developing co-created and co-supervised internships of summer students between a biophysicist researcher and a business management academic with a remit to work up a suite of useful documents and processes and recommend productivity management tools.
- *Can innovative career paths forged from blending of academic and industry support, including substantive industrial buy-out and secondment of postdoctoral researchers and technicians, catalyse industrial impact of discovery-led research outcomes?*
    - If there were a pool of research specialists or research officers, a university could perhaps support a proto-startup by providing an in-kind donation of, for example, ca.

30% of a research officer specialist's support time for the first 2 years of the company's life, tapered out subsequently to help nurture sustainability. Universities could utilise these research officer specialists in a coordinated way as a 'pool of experts' ready to turn their hand to a vast range of new industry-facing problems. There is precedence in the form of research software engineers (RSEs). RSEs are experts not only in software development and good practice, but can also write specialised machine control software, and help develop experimental protocols and pipelines – for example, tackling challenging questions such as how best to make and run a machine that does buffer exchange, as well as interfacing with hardware such as bespoke microscopy and robotics. Similar banks of biophysics-cognisant research experts in microscopy, biochemistry, nanofabrication, and so forth, may result in myriad benefits to incipient biophysics commercialisation opportunities. Credible business models cloning past resource frameworks for RSEs might capture a relevant portion of a research officer's time onto research grant application. However, since there is uncertainty of success for any grant universities would need to front-load investment in these roles which comprises risk. A key is that income is used to subsides a pool of open-contracted expertise to decouple any individual employment contract from the fixed term duration of any individual research grant but rather create genuinely openly contracted posts that encourage the retention of this significant expertise. The opportunities offered by such an expert interdisciplinary research support team are significant. Such teams could collaborate across multiple departments and research groups as well as industry partners. To help retain the best expertise might require capturing innovative career development opportunities for these specialists such as permitting a portion of their time, for example ca. 20%, to be dedicated to developing their own independent research. At a practical level, such research whether dedicated to their own personal aspirations or in support of other research teams, is likely to be most beneficial if delivered in unitary blocks of time which reflect the timescale of supporting innovative biophysics experiments and analysis – for example weeks as opposed to hours or days.

- *Despite the concomitant risk, is there a case for more aggressive University seed investment towards promoting commercialised innovation from academic research?*
    - Most universities are fortunate in that they typically employ dedicated personnel to advise on commercialisation opportunities. However, it may be possible to achieve more through greater investment into project managers who interface between academia and industry. For example, to follow a biophysics research officers business model, can academic institutions also support the development of pools of delocalised expertise in project management? Such individuals could potentially support the operational activities of several biophysics research projects which are often complex require high levels of project coordination across multiple research teams, including industrial partners.
- *Could UK plc be benefiting more from the substantial expertise generated through PhD and postdoctoral biophysics research?*
    - The way the UK has traditionally trained its junior researchers in biophysics has been largely focused on the researcher engaging just with one, or a small handful, of academic institutions, in relatively niche biophysics techniques associated with often very focused granular scientific questions. A subset of such students and postdocs may potentially engage with commercialisation opportunities through industry partnerships associated either with specific research projects or wider doctoral training centre initiatives in the case of PhD students. However, there are alternative international models enabling more *delocalised* junior researcher training and better engagement with industry partnerships.

An example has been a European Commission funded MSCA Doctoral Networks (DN) 11, formerly known as Innovative Training Networks (ITNs), which developed from the ITN scheme within Horizon 2020 and is now part of Horizon Europe. DNs are European-level training initiatives that fund doctoral programs through international networks of academic and non-academic institutions, with an aim to train a new generation of "creative, entrepreneurial, and innovative researchers, equipping them with the skills needed for international and interdisciplinary collaboration." In practice, a network comprises ~10 academic and ~5 non-academic (primarily industrial) partners, with 10-15 PhD students (typically administratively homed with one of the academic partners. Whilst receiving granular local training relevant to their individual projects from their home research team the whole cohort will typically *en masse* gain collective training every few months at one or more of the other partners, including the industrial nodes of the network. Such training is also supplemented by more focussed secondments for each student between different partners throughout the duration of their PhD. Such a flexible framework enables a far greater breadth of training, adds value through cohort and peer-peer training with researchers across multiple tiers of disciplinary expertise engaging each other in common scientific discussion, and critically involves both academic and a wide range of non-academic partners focused on industrial applications and commercialisation potential. An interesting future possibility which may be beneficial to explore could be to widen out such a network of partner expertise to the whole of the UK. Namely, to create a national, coordinated junior researcher training capability in biophysics. In fact, the successes of the Physics of Life network (PoLNET)[15] may have established the key foundations in the architectures needed to support such an innovative training approach, and one which potentially has the ability to extend to postdoctoral as well as graduate student training, including not only direct training activities but career development opportunities such as delocalised *mentorship* on a national scale.

- *Can more be done to educate UK policymakers and the public of the need for basic science to bring about innovation?*
  - The level of perception from British biophysics researchers that basic science is the key driving force for innovation is high, as is the evidence discussed above. However, politicians are often naïve about how science works, and impatient due to too great an emphasis on short-term electoral cycles. Even Margaret Thatcher (UK Prime Minister 1979-90, who started graduate research in Chemistry/Biophysics investigating aspects of lipid bilayers), almost decimated the science base in the UK when serving as serving as the Secretary of State for Education and Science 1970-74. Historically, right wing politicians in the UK have tended to support projects and hardware, whereas left wing politicians have supported labour and people. A challenge in moving forwards to have a balance of both. One route forward towards maintaining an appropriate level of science education in UK policymakers is through the Parliamentary Office of Science and Technology (POST) – an impartial research and knowledge exchange service based in the UK Parliament that publishes independent, topical briefings, drawn from research evidence, and tailored to the needs of parliament[16]. Ultimately, politicians serve the wider public however, and there is a clear need to for scientists to continue to engage the public directly on how basic science drives innovation.
- *Do we need to be more proactive in addressing the risks of losing good UK biophysics startups due to the pressures of unconstrained market forces and globalisation?*
  - An "Invisible Hand" in economics, as Adam Smith first coined the metaphor[17], suggests that individual self-interest (i.e. making a profit) may inadvertently contribute to an overall wider well-being, such as at the level of society. However, this mantra of modern capitalisation can result in undesirable outcomes when there is a tension of profit versus societal benefit, as arguably demonstrated in the UK minimising Personal Protective Equipment (PPE) storage prior to the COVID pandemic, and also in the incentive for owners to maximise individual wealth by selling their shares in UK startups to global corporations who are not fully invested in their future development. It is difficult to predict how popular advocating overly protectionist economic policies would be, especially considering the substantial current fluidity in global economics impacted by recent changes in US import tariffs legislation. However, one idea which could have more tangible merit would be for UKRI, UK Government and research-intensive UK universities to invest together into a common budget, which could then invest in startups under expert guidance. Such a *mutual fund* might ensure that UK interests can be spoken for in future proposed selloffs of biophysics startups, as well as helping to inform several other high-level strategic decisions, bringing a share of revenue back for reinvestment, and ultimately making this process self-sufficient and non-political; arguably more focused around the broader interests of the UK taxpayer. The shared risks of such a mutual fund might also allow more speculative investment and reduce the burden on individual academic institutions. Conversely, biophysics researchers and associated startups might also be more agile and less constrained by local institutional needs. With the caveat relating to current global economic fluidity that has been an impact of the recent US import tariffs legislation, the UK currently exports a significant amount of scientific instrumentation abroad – equating to a ca. 45% increase over the period 2018-23 with an equivalent value as of 2023 of almost £3 billion[18]; with current high energy costs, and high labour costs, the UK risks losing even this position without proactive measures:
*here is urgency to retain scientific instrumentation development capability and the driving ideas and expertise behind this in the UK.*

**Discussion and Conclusions**

UK biophysics has for several decades generated success stories in discovery-led research that brews invention but has traditionally lagged behind international counterparts in regard to subsequent innovation and sustainable commercialisation. The UK punches above its weight in terms of developing new biophysics technology, but there is scope to improve its structures to help facilitate commercialisation; we could punch even higher given the right approach by UK Government. For example, according to a recent 2024 report, the $25.4 billion budget of the National Aeronautics and Space Administration (NASA) in the US generated $75.6 billion in economic output[19], suggesting an equivalent "multiplier" of research income to output of ca. 3. Obtaining equivalent precise figures for UK biophysics innovation is non-trivial, however it is almost certainly much lower than this, demonstrating the power of significant investment at a state sponsored level.

What we have tried to do in this review is firstly reflect constructively on the historical narratives from a handful of emerging success stories in which biophysics science researchers have become genuine entrepreneurs but also tension these against lessons which can be learnt from examples of failed commercialisation in the UK. We have discussed how structural revisions to the way we sustainability resource researchers in senior biophysics postdocs could be beneficial, as could in-house management training for academics coupled with additional support from expert project managers to help navigate the complex interactions that are critical to properly negotiating deals with industrial partners and investors, and for the efficient running of complex biophysics research project which have embedded commercialisation opportunities. We have also discussed possibilities for restructuring the way we train and mentor our national pool of junior researchers in biophysics to best tap into to the enormous breadth of biophysics expertise within the UK.

Such initiatives all have resourcing implications, however, and may be difficult to achieve in times of financial uncertainty across the higher education sector in the UK, as well as potentially falling foul of administrative barriers when attempting to reach agreement involving multiple academic institutions. However, there is one barrier to future biophysics commercialisation success which the researcher community is empowered to resolve autonomously, which is cultural. This works on two levels. Firstly, in UK STEM research there is still not, as a rule, a strong cultural expectation that a focused discovery-driven research project should necessarily lead to, or at least help facilitate, commercialisation. Conversely, other countries have different expectations, for example in Nordic countries where ultimate commercialisation is a more common expectation. Secondly, there is the academic-island effect, of the emergence of mini-fiefdoms within research that ultimately serve to create more academic silos as opposed to building collaborative bridges, sharing of research resources and expertise. Such features of research culture can be addressed and modified through open and collegiate discussion and a willingness to together work as a *national community* of biophysics researchers. The resources required to catalyse such discussions are arguably minimal, but the positive impact on the future of UK plc and the quality of life of its residents could be enormous.

## Acknowledgements

Supported by funding from the Engineering and Physical Sciences Research Council (EPSRC, grant references EP/T022000/1, EP/W024063/1 and EP/Y000501/1). Many thanks to the biophysics entrepreneur contributors to the Physics of Life 2025 roadmap who gave their permission for their testimonials to be used.

## References

1. Physics of Life Roadmap 2025. https://www.physicsoflife.org.uk/physics-of-life-roadmap.html.
2. Li, Z., Gothard, E., Coles, M. C. & Ambler, C. A. Quantitative Methods for Measuring Repair Rates and Innate-Immune Cell Responses in Wounded Mouse Skin. *Front Immunol* **9**, (2018).
3. Chisholm, D. R. *et al.* Photoactivated cell-killing involving a low molecular weight, donor-acceptor diphenylacetylene. *Chem Sci* **10**, 4673–4683 (2019).
4. Maglia, G., Restrepo, M. R., Mikhailova, E. & Bayley, H. Enhanced translocation of single DNA molecules through α-hemolysin nanopores by manipulation of internal charge. *Proc Natl Acad Sci U S A* **105**, 19720–19725 (2008).
5. Craggs, T. D. & Kapanidis, A. N. Six steps closer to FRET-driven structural biology. *Nature Methods 2012 9:12* **9**, 1157–1158 (2012).
6. Hohlbein, J. *et al.* Conformational landscapes of DNA polymerase I and mutator derivatives establish fidelity checkpoints for nucleotide insertion. *Nature Communications 2013 4:1* **4**, 1–11 (2013).
7. Ambrose, B. *et al.* The smfBox is an open-source platform for single-molecule FRET. *Nature Communications 2020 11:1* **11**, 1–6 (2020).
8. Kapanidis, A. N. *et al.* Initial Transcription by RNA Polymerase Proceeds Through a DNA-Scrunching Mechanism. *Science (1979)* **314**, (2006).
9. Li, H., Ying, L., Green, J. J., Balasubramanian, S. & Klenerman, D. Ultrasensitive coincidence fluorescence detection of single DNA molecules. *Anal Chem* **75**, 1664–1670 (2003).

10. Andrecka, J. *et al.* Structural dynamics of myosin 5 during processive motion revealed by interferometric scattering microscopy. *Elife* **4**, (2015).
11. Young, G. *et al.* Quantitative mass imaging of single biological macromolecules. *Science (1979)* **360**, 423–427 (2018).
12. Yan, H. *et al.* Thermo-reversible protein fibrillar hydrogels as cell scaffolds. *Faraday Discuss* **139**, 71–84 (2008).
13. Regional research capacity: what role in levelling up? - HEPI. https://www.hepi.ac.uk/2024/01/18/regional-research-capacity-what-role-in-levelling-up/.
14. Universities for North East England launch - Durham University. https://www.durham.ac.uk/news-events/latest-news/2024/11/universities-for-north-east-england-launch/.
15. UK Physics of Life (POLNET). https://www.physicsoflife.org.uk/.
16. About us - POST. https://post.parliament.uk/about-us/.
17. Book IV, Chapter 2 | Adam Smith Works. https://www.adamsmithworks.org/documents/book-iv-chapter-2.
18. UK trade with the United States - Office for National Statistics. https://www.ons.gov.uk/economy/nationalaccounts/balanceofpayments/articles/uktradewiththeunitedstates2023/2023#uk-trade-in-goods-with-the-united-states-interactive-content.
19. NASA Economic Impact Report 2024. https://space-economy.esa.int/article/216/nasa-economic-impact-report-2024.